\documentclass[12pt]{article}
\usepackage{amsbsy}
\newcommand{\nc}[2]{\newcommand{#1}{#2}}
\newcommand{\ncx}[3]{\newcommand{#1}[#2]{#3}}
\nc{\ba}{\begin{array}}
\nc{\ea}{\end{array}}
\nc{\bab}{\left(\begin{array}} 
\nc{\eab}{\end{array}\right)} 
\nc{\bea}{\begin{eqnarray}}
\ncx{\beal}{1}{\begin{eqnarray}\label{#1}}
\nc{\eea}{\end{eqnarray}}
\nc{\bean}{\begin{eqnarray*}}
\nc{\eean}{\end{eqnarray*}}
\nc{\be}{\begin{equation}}
\ncx{\bel}{1}{\begin{equation}\label{#1}}
\nc{\ee}{\end{equation}}
\nc{\nit}{\noindent}
\nc{\ra}{\rightarrow}
\nc{\del}{\delta}
\nc{\ev}{\equiv}
\nc{\ito}{\infty}
\nc{\lb}{\left(}
\nc{\lsb}{\left[}
\nc{\ld}{\left.}
\nc{\rb}{\right)}
\nc{\rsb}{\right]}
\nc{\rd}{\right.}
\nc{\e}{{\rm e}}
\nc{\dr}{{\rm d}}
\nc{\eln}{\nonumber\\[2mm]}
\nc{\el}{\\[2mm]}
\nc{\da}{&=&}
\nc{\ii}{{\rm i}}
\nc{\eps}{\epsilon}
\nc{\og}{\omega}
\nc{\rg}{\rho}
\nc{\gam}{\gamma}
\nc{\grad}{\vec\nabla}
\nc{\pl}{\partial}
\ncx{\rf}{1}{(\ref{#1})}
\ncx{\avec}{1}{\stackrel{\leftarrow}{#1}}
\ncx{\der}{2}{\frac{\dr #1}{\dr #2}}
\begin{document}
\title{\bf The Lorentz Group, a Galilean Approach.}
\author{D. E. Jaramillo and N. Vanegas. \\
{\small \em Instituto de F\'\i sica, Universidad de Antioquia},\\ 
{\small \em A.A. 1226, Medell\'\i n, Colombia}.} 
\maketitle 
{\footnotesize 
\nit We present a  pedagogical approach to the Lorentz group. We start
by introducing a compact notation to express the elements of 
the fundamental representation of the rotations group.    Lorentz
coordinate  transformations are derived in a novel and compact form.
We show how to make  a Lorentz transformation on the
electromagnetic fields as well. A covariant time-derivative is introduced in 
order to deal with non-inertial systems. Examples of the usefulness of these results
such as the rotating system and the Thomas precession, are also presented.

\section{Introduction} 
Special relativity was first introduced nearly  a century ago in order to
explain the massive experimental evidence against  ether as the medium for
propagating electromagnetic  waves. As a consequence  of special relativity an
unexpected space-time structure was discovered. The pure Lorentz 
transformations called  boosts relate the changes of the space distances and 
time  intervals  when they  are measured from  two different inertial frames.  
Rotations and boost transformations form the  general Lorentz group
(The properties of the Lorentz group can be found in other references
such as \cite{1a}-\cite{2a}).  

We show how one can understand  boost
transformations,  which follow from the postulates of  special relativity, 
as corresponding to   deformations of the classical Galilean transformations. 
Also we  introduce  a covariant temporal derivative to deal with non-inertial
systems. 
This article is arranged as follows.
In section II we show a simple way to  generate and write the matrices 
associated with  the  rotation of three dimensional vectors 
and  present some applications of our notation. 
In section III we   find   the matrices of the  boost  transformations starting
from  Galileo's  only by imposing the constance of the velocity of  light.
Finally, in section IV we show how the electromagnetic fields  transform under
general Lorentz transformations in the same fashion we introduced before.
An appendix deal with non-inertial  system. 
\section{Rotations}
\subsection{Rotations of the Coordinate Frame}\label{section:rot}
 Under rotations the Cartesian coordinates of  a specific
vector transform linearly  according to
\bel{r1} 
\vec x\ra \vec x'= R \vec x,
\ee 
so that  
\bel{r2}
\vec x\cdot\vec x=\vec x'\cdot \vec x'.
\ee 
In a three dimensional space,   $R$  corresponds to
a  $3\times3$ orthogonal  matrix and the array $\vec x$  is written as a
column. In order to find explicitly the $R$ matrix we   analyze 
infinitesimal rotations  and, as usual, then construct a finite transformation,
 made of an  infinite number of infinitesimal ones. If an
infinitesimal transformation is represented by   
\bel{r3} \vec x\ra \vec x'= 
        \vec x-\del \vec x , \ee   
then, from \rf{r2}, $\del \vec x$  in first
approximation satisfies  
        \bel{r3a}\del \vec x \cdot\vec x=0\ee
for all $\vec x$. 
The solution of this equation is given by
        \bel{r4}\del \vec x=\del\vec  \theta \times
                \vec x,\ee 
the infinitesimal vector $\del\vec \theta$ physically carries the total
information about of the rotation: $|\del\vec \theta|$  gives the
magnitude of the  rotation angle and    
 $ \hat{\theta}\ev\del\vec \theta/|\del\vec \theta|$  are the coordinates of 
 the unit vector,  parallel to the rotation axis.
From this the (infinitesimally) transformated coordinates are  written  as
        \bel{r5}
                \vec x'=(1-\del\vec\theta\times)\vec x .\ee
The expression in brackets corresponds to the infinitesimal rotation
matrix $R(\del\vec \theta)$.  
The quantity $\del\vec \theta\times$  is a  (matrix) operator 
which can be defined as follows
\bel{r7}
(\del\vec\theta\times)\vec x\ev \del\vec \theta\times\vec x,
\ee
or more explicitly,
\bel{r8}
        \del\vec \theta\times=\bab{rrr} 0\;&-\del\theta_3&\del\theta_2\\
        \del\theta_3&0\;&-\del\theta_1 \\
        -\del\theta_2&\del\theta_1&0\;\eab. \ee

Writting 
        \[ \del \vec \theta =\lim_{N\ra\ito} \vec\theta/N\]
the  matrix for a finite angle $\vec\theta$ rotation corresponds to 
\bel{r6} \label{eq:covariant-deriv}
        R(\vec\theta)=\lim_{N\ra\infty}\Big[R(\vec\theta/N)\Big]^N=
        \lim_{N\ra\infty}\lb1-\frac{\vec\theta\times}{N}\rb^N= 
        \e^{-\vec\theta\times}.\ee
The  expansion of the exponential  in
\rf{r6} gives us   the $R$ matrix explicitly,

\bel{r10} \e^{-\vec\theta\times}=  \hat{\theta} \hat{\theta} \cdot - 
        \sin \theta\; \hat{\theta}\times  - \cos \theta (\hat{\theta}\times)^2, 
        \ee 
which applied to the coordinates
gives the conventional expression of  coordinate rotations \cite{2a}.
For arriving to \rf{r10} we have used  the properties of the triple 
vector product to obtain
\bel{r9}
(\vec\theta \times) (\vec\phi \times)= \vec\theta \vec\phi\cdot - \vec\theta
\cdot\vec\phi; \ee the last term is understood to be the coefficient of an
identity matrix.  In this notation the period after a vector implies  its
transposition:  $ \vec \theta\cdot\ev \vec \theta^T$. 

\subsection{Rotations Algebra.}
\label{section:rot-alg}
As is well known a  group is a set of operators with a multiplication
law which satisfies
four basic properties: closure, associativity, existence of the identity and
the existence of a unique inverse for each element.  The set of rotation
matrices $R$ represents a group: the rotation group. The elements of the rotation
group are labeled by the set of continuos parameters $ \theta_i$. The
antisymmetric matrix $\vec\theta \times$ generates the rotation matrix
$R(\vec\theta)$, this is why it is called ``generator". Generators form a
vector space as well.  The rotations algebra is the commutation relations
among the elements of the generators vector space basis.

The  closure property it is nothing  more than the statement that
 the composition of  two rotations is again a
rotation. This is implemented in group theory language by saying  that   the
commutator  between two generators is a generator.
For the generators of the rotation group we obtain
\bel{so1}  [\vec \theta\times,\vec \phi \times] =(\vec
\theta\times\vec \phi)\times , \ee  
where we have used  the Jacobi identity for the triple vector product.

If  the $\hat{ e}_i$ form the standard basis of the coordinate
space, they satisfy  the algebra
\bel{so2}\hat{ e}_i\cdot \hat{ e}_j=\del_{ij},\;\;\;\hat{ e}_i
\times\hat{ e}_j =\eps_{ijk} \hat{ e}_k,
 \ee
where $\eps_{ijk}$ is the
totally antisymmetric Levi-Civita tensor. (The sum over the repeated  indexes
is understood.) 
Writing

\bel{so3} {\cal J}_i= \ii \hat{ e}_i\times\ee
we find that the generators can be re-written as   
\be \vec \theta\times=- \ii \vec \theta \cdot\vec {\cal J};\ee
that is, $\vec {\cal J}$ corresponds to  a hermitian base for the generator 
space. According to \rf{so1} and
\rf{so2}   the ${\cal J}$'s then satisfy

\bel{so4}
[{\cal J}_i, {\cal J}_j]=\ii \eps_{ijk} {\cal J}_k .
\ee

The relation \rf{so4} corresponds to the algebra of rotations.

\nc{\dl}{In that case the generators act over  the three dimensional representation (the definitory one)
We can write these generators acting over  one infinite dimensional representation into  the Hilbert
space.
We espect that the scalar wave funtion $\psi(\vec x)$ under rotations transforms like
\[  \psi'(\vec x')= \psi(\vec x)  \]
where $\vec x'$ and $\vec x$ are related by a rotations transformation \rf{r1}
so  we can write
\[  \psi'(\vec x)= \psi( R^{-1}\vec x)  \]
The infinitesimal transformation is
 \[  \psi'(\vec x)= \psi(\vec x- \del\vec x )
 =\Big(1- \del\vec x\cdot \grad \Big) \psi(\vec x)  \]
 which according to \rf{r4} can be written
\[  \psi'(\vec x)= \Big(1- \del\vec\theta\times \vec x\cdot\grad\Big) \psi(\vec x)
= \Big(1- \del\vec\theta\cdot \vec x\times\grad\Big) \psi(\vec x) \]
Now compouning a finite rotations by adding infinite infinitesimal rotations we obtain

\[  \psi'(\vec x)=\e^{-\vec\theta \cdot \vec x\times\grad}   \psi(\vec x)\]

\[ -\vec\theta\cdot \vec x\times\grad \ev -\ii \vec \theta\cdot \vec L\]
In this case the generators of the rotations are represented by the operartor
\[  \vec L= -\ii\vec x\times\grad \]
and the components are
\[ L_i=\hat e_i \cdot \vec L= \vec x\cdot (\ii\hat e_i  \times)\grad
= \vec x\cdot{\cal J}_i\grad \]
The commutation relation among the $L$'s are
\[ [L_i,L_j]=  \vec x\cdot[{\cal J}_i,{\cal J}_j]\grad= \ii\eps_{ijk} L_k  \]
where we have used the fact that
\[ \grad  \vec x\cdot  =1\!\!1\]
(the quadratic derivatives dissapear because they are symmetric in the $i,j$ indexes and
the commutator choice the antisymmetric part)}

\subsection{Rotating Systems} \label{section:rot-sys}
All of the subsection [\ref{section:rot-alg}] is standard, however in connection with
subsection [\ref{section:rot}] we can obtain interesting results.  As an example of the
usefulness of the notation introduced in \rf{r6} for the rotation matrix,
let's find the velocity and acceleration of a particle observed from a
rotating system. Let a vector $\vec x$ be the coordinates of a
particle in an inertial system and $\vec x'$ the coordinates of the same
particle observed from a rotating system, with angular velocity
$\vec \og$; the origins of these two systems are located at the same
geometrical point so that the coordinates satisfy the relation

\bel{rs1} \vec x'=\e^{-\vec \theta\times}\vec x ,\ee
where $\vec \theta$ is a time-dependent function.
In the inertial system the velocity and acceleration of one particle  are the
first and second time-derivative of the coordinates, respectively. 
Assuming that the components of a force, acting over the particle, transform
according to \rf{rs1} we conclude that, in the rotating system, the second
Newton law $\vec F= m\vec a$ does not have this form, unless we change the
time-derivative to  a covariant time-derivative given by
\bel{rs2} 
        D_t\ev \e^{-\vec \theta \times}\der{}{t}\e^{\vec \theta\times}
        =\der{}{t}+\vec \og \times+\frac{1}{2}(\vec \og \times\vec\theta)\times
        +\frac{1}{3!}\Big((\vec \og \times\vec\theta)\times\vec\theta\Big)\times
 +...\ee
where we have used \rf{so1} in the known relation
\[ \e^{-A}B\e^{A}=B+[B,A]+\frac{1}{2} [[B,A],A]+\frac{1}{3!}
        [[[B,A],A]],A]+...\]
Thus we can define a covariant  velocity $\vec v'$ of the particle, seen in 
the rotating system, as the covariant derivative of the coordinates;
%
in the simple case   in which $\vec \omega$ is paralell to
$\vec\theta$ we have  

\bel{rs3}
 \vec v'= \der{\vec x'}{t}+\vec \og\times \vec x' .\ee
In the same way the covariant acceleration is then given by   

\bel{rs4}
 \vec a'=D_t{\vec v'}= \der{^2\vec x'}{t^2}
+2\vec\og\times\der{\vec x'}{t}+ \vec\og\times(\vec\og\times\vec x') +\vec
\alpha\times\vec x'.
\ee
where $\vec\alpha$ is the angular acceleration of the system. 
In the second term of the RHS  we recognize the Coriolis
acceleration \cite{2a}-\cite{3c}, 
and   the centrifugal acceleration in the third term.
In this way  the primed vectors are  related  with the un-primed
quantities by a relation similar to \rf{rs1}.

\section{Lorentz Transformations.}

Lorentz transformations are the rules that relate space-time coordinates of
any event in two different inertial systems. Basically, Lorentz
transformations can be classified in two types, rotations and boosts. A
general Lorentz transformation is a mixing between them.  Boosts are the
Lorentz transformations when the systems have parallel spatial axis with
spatial origin in relative movement. As we will see, Lorentz
transformations are the generalization of the classical rotations to
4-dimensional space-time.

\subsection{Boost Transformations.}

In order to deduced how to transform the coordinates of any event after a
boost let us take $S'$ to be an inertial system in relative movement with
respect to another inertial system $S$.  The respective axes in both
systems are parallel. Take also their spatial origin as coincident at time
zero for both systems. We get that the space-time origin of the two systems
is the same.  According to the Galilean transformations, in that case, the
coordinates $t'$ and $\vec x'$ of a event,as observed from $S'$, are
related with the $t$ and $\vec x$ coordinates of $S$ given by \beal{lt1}
t'\da t\eln \vec x'\da \vec x- \vec u t ; \eea where $\vec u$ is the
velocity of $S'$ relative to $S$.  As a result of these relations the
velocity of one particle observed in $S'$ is the velocity observed by $S$
minus the relative velocity $\vec u$.  Clearly this is in contradiction
with the postulate of special relativity that the speed of the light is
constant independently of the choice of coordinates, because that relation
of velocities remains true even when a light pulse is considered instead of
a particle.

According to the special relativity principles if we  suppose that a light 
pulse is emitted from the origin  the  space-time coordinates,  the pulse must
satisfy  

\bel{lt2}
\vec c^2t^2-\vec x^2=c^2t'^2-\vec x'^2=0.
\ee

One can, however, try to modify the Galilean transformations to make it
compatible with the relativity principles, let us proceed like this; for
the $u/c\ra 0$ approximation take the deformed Galilean transformations to
be

\beal{lt3} t'\da t-\del t\eln \vec x'\da \vec x- \vec u t,\eea 
introducing a factor $\del t$. In order to satisfy  \rf{lt2} 
in first approximation we obtain 

\bel{lt4}\del t= \frac{\vec u\cdot\vec x}{c^2}.\ee
Notice  that \rf{lt2} together with \rf{lt3}  satisfy  the first equation in
\rf{lt1}   even  if  $c^2t^2-\vec x^2$ vanishes. That is, even if $\vec
x$  and $t$ represent the coordinates of any arbitrary event.
These  so deformed Galilean transformations correspond to infinitesimal
boost transformations.

It is convenient to define a infinitesimal parameter  as
\bel{lt5}\del\vec  \eta \ev \ld \frac{\vec u}{c}\right|_{u/c\ra0} .\ee
We can  write   the infinitesimal Lorentz transformation \rf{lt3}, 
 using  \rf{lt4} and \rf{lt5},  as the following  matrix equation

\bel{lt6} \bab{c} ct'\\ \vec x'\eab =\lsb  
1-\bab{cc} 0&\del\vec\eta\cdot\\ \del\vec\eta &0 \eab 
\rsb \bab{c} ct\\ \vec x\eab .
\ee
Assuming  
\[
\del\vec\eta=\lim_{N\ra \infty}\vec\eta/N,
\]
one  can reconstruct the
finite Lorentz transformations, using a procedure similar to the one
introduced in (\ref{eq:covariant-deriv}); performing an infinite number of
infinitesimal transformations the result is 
\bel{lt7}
\bab{c} ct'\\ \vec x'\eab =
\lim_{N\ra \infty}\lsb  
1-\frac{1}{N}\bab{cc} 0&\vec\eta\cdot\\ \vec\eta &0 \eab 
\rsb ^N \bab{c} ct\\ \vec x\eab =
\exp \bab{cc} 0&-\vec \eta \cdot\\ -\vec \eta &0 \eab  \bab{c} ct\\ \vec x\eab .
\ee
Expanding the exponential we obtain
\bel{lt8}\exp \bab{cc} 0&-\vec\eta \cdot\\ -\vec\eta &0 \eab 
=\bab{cc} \cosh\eta&-\hat{ u}\cdot\sinh \eta\\- \hat{ u} 
\sinh \eta& \hat{ u}\hat{ u}\cdot\cosh\eta-(\hat{ u}\times)^2 \eab .
\ee
From \rf{lt7} and \rf{lt8} we can work out the relative velocity between
the two 
coordinate systems  
\bel{lt9}\vec u=-\ld\frac{\vec x' }{t'}\right|_{\vec x=\vec 0}=\hat{
u}\tanh\eta,\ee  therefore
\bel{lt10}
\sinh\eta=\frac{u/c}{\sqrt{1-u^2/c^2}} ;\;\;\cosh\eta= 
\frac{1}{\sqrt{1-u^2/c^2}}\ev \gam.
\ee
Thus \rf{lt9} gives the  relation between the parameter $\eta$ and the
relative velocity $u$. It is evident that if $u/c\ra 0$ we get  $\eta\ra
u/c$; for this reason $\eta$ is called the relative ``rapidity".
 %
%

In general, a Lorentz vector is a 4-vector which transforms according to
\rf{lt7} (with \rf{lt8} and \rf{lt10}). Just by introducing a deformation
to the Galilean transformations one can introduce the results of special
relativity and motivate the necessity of a constant speed of light (for any
observer). 

\subsection{Lorentz  Algebra.}
As in section [\ref{section:rot-alg}] once we know the way a vector
transforms we can find out about the group algebra that these
transformations imply.  From the expression \rf{lt7} one  can
guess the generators of a boost 
transformation. The set of  boost transformations does  not form a
group, this can be  seeing  by the fact that    the commutation relation
between  boost generators is not a  boost generator itself,  
\bel{s1}
\lsb\bab{cc} 0&\vec\eta \cdot\\ \vec\eta &0 \eab,
\bab{cc} 0&\vec\kappa \cdot\\ \vec\kappa &0 \eab\rsb =
\bab{cc} 0&\vec 0\cdot\\ \vec0&-(\vec\eta\times \vec\kappa)\times  \eab.
\ee
Nevertheless this generators form a vector space which can be expanded in 
the  basis of $\vec {\cal K}$,  defined by
\bel{s2}
{\cal K}_i=\bab{cc} 0&\hat{ e}_i \cdot\\ \hat{ e}_i&0 \eab.
\ee
The commutation relations \rf{s1} for the $\cal K$'s are 

\bel{s3}
[{\cal K}_i,{\cal K}_j]=\ii \eps_{ijk} {\cal J}_k,
\ee
where, in this case, the $\cal J$'s  are the  rotation generators given in
\rf{so2}  extended to four dimensions,

\bel{s4}
{\cal J}_i=\bab{cc} 0&\vec 0 \cdot\\ \vec 0&\ii\hat{ e}_i\times \eab.
\ee
The generators  $\cal K$ do not form a closed  algebra,  
$\cal K \oplus J$'s do, the algebra closes with

\bel{s5}
[{\cal J}_i,{\cal K}_j]=\ii \eps_{ijk} {\cal K}_k .
\ee
Relations \rf{so4}, \rf{s4} and \rf{s5} form the
Lorentz Algebra. This algebra is a manifestation  of  the fact that  rotations,
together with  boosts, form a group,  the Lorentz group. The $\cal
K$'s and $\cal J$'s are a basis for  the generator space of this group.
We can change the basis, in particular a good choice is the basis compounded by
the $\cal N$'s and their complex conjugate $\cal N^*$'s  defined by  
 \bel{s6} 
 {\cal N}_i \ev {\cal J}_i +{\cal K}_i,
\ee
which satisfy the algebra 
\bel{s7}
[{\cal N}_i ,{\cal N}_j ]=2\ii \eps_{ijk}{\cal N}_k, 
\ee
that is $\frac{1}{2}{\cal N}_i$ and $-\frac{1}{2}{\cal N}_i^*$ satisfy
independetly satisfy the rotation algebra \rf{so4}, additionally,
\bel{s8}
[{\cal N}_i,{\cal N}_j^*]=0.
\ee
We see that the Lorentz algebra  can be splitted into two ``rotation" invariant
subalgebras. 

\subsection{Thomas precession}

 Relation \rf{s3} correspond to the application of two consecutive boosts;
 it shows that a vector is rotated when these two boosts are applied.
 This phenomena is known as the Thomas
precession. Physically the  Thomas precession   appears when we try to
describe the time evolution of quantities asociated to accelerated particles.

In order to analyze the problem of an accelerated particle, the usual
thinking is of a non-inertial system as composed of infinite inertial
system where the particle is always instantaneously at rest in one of them
\cite{1a}, \cite{4a}, \cite{4b}. However, as we will see, this problem is equivalent (at least
locally) to considering only one non-inertial rest frame where the
``boost'' from the  laboratory system is characterized by a time depending
rapidity $\vec\eta(t)$.

For the non-accelerated particle the time derivative used in the
laboratory system changes as
\[
 \der{}{t}\ra
\der{}{t'}=\frac{1}{\gam}\der{}{t},
\]
when the  observer  uses    the  system where the particle is at rest.
 
Following the procedure of section [\ref{section:rot-sys}], for an
accelerated particle, we must define a covariant time derivative for an
observer in the frame in which the particle is at rest, as with the
the rotating system \rf{rs2},
                                                                              
\bel{tp1} \der{}{t}\ra D_{t}=
\e^{-\vec \eta\cdot\vec {\cal K}} \der {}{t'} \e^{\vec \eta\cdot\vec {\cal K}}
\ee
In the non-relativistic approximation,  and considering \rf{tp1}
acting only on 3-vectors (see appendix)  we have 
 \bel{tp2}
D_{t}=\der{}{t}+\lb\frac{\vec u \times\dot{ \vec u}}{2c^2}\rb\times 
\ee
where $\vec u$ is the velocity the particle seen from the laboratory system. 
Comparing with \rf{rs2} we find that this system has a precession
frecuency given by \bel{tp3}
\vec \og=\frac{\vec u \times\dot{ \vec u}}{2c^2}\ev -\vec\og_T
\ee
$\vec\og_T$ is called Thomas frecuency.
For instance, the time evolution of the spin vector of a accelerated particle
of mass $m$, charge $e$  and gyro-magnetic ratio $g$ is not 
$\dr{\vec s}/\dr{t}= g(e/ {2m})\, \vec s\times \vec B'$
but 
\bel{tp4}
\der{\vec s}{t}-\vec \og_T\times \vec s=
g\frac{e}{2m}\vec s\times \vec B'
\ee 
where $\vec B'$ is the magnetic field observed in the rest frame of the
particle. Once again, following the method introduced in classical
mechanics and deforming the Galilean set of transformations one is able to
obtain, without too much effort, a fundamental result of relativistic
mechanics. 

\section{Transformations of the Electromagnetic Field} 

In the same spirit  of this paper,  Maxwell
equations with sources can be written in a matricial form as  
\bel{em1}
\bab{rc} 0& -\vec E\cdot\el -\vec E&\vec B\times\eab 
\bab{c} \avec \pl_0 \el -\avec \nabla\eab=\bab{c} \rg \el \vec J\eab,
\ee
where $\avec~$ over the derivatives means that they  act to the right. We are
assuming $c=\eps_0=1$ for simplicity. (Homogeneous Maxwell equations are
obtained by duality, $\vec E\ra -\vec B$, $\vec B\ra\vec E$, $\rho\ra0$.) 
We can then write the electromagnetic field array as a combination of the
generators of the Lorentz group; in our notation  
 \bel{em2} 
\bab{rc} 0& -\vec
E\cdot\el -\vec E&\vec B\times\eab=-(\vec E\cdot\vec { \cal K}+\ii \vec
B\cdot\vec{ \cal J})  
\ee 
Under Lorentz transformations  the spacetime
derivative and the sources in \rf{em1} transforms like the coordinates in
\rf{lt7},  so the  matrix of the electromagnetic fields transform according to
  \bel{em3}
\vec E'\cdot\vec {\cal K}+\ii \vec
B'\cdot\vec{ \cal J}= \e^{-\vec \eta\cdot\vec { \cal  K}}(\vec E\cdot\vec {  \cal K}+\ii \vec
B\cdot\vec{  \cal  J}) \e^{\vec \eta\cdot\vec {  \cal K}};
\ee
taking infinitesimal transformations for the fields we find
\[
\vec E'\cdot\vec {  \cal K}+\ii \vec
B'\cdot\vec{ \cal J}=\vec E\cdot\vec {\cal  K}+\ii \vec
B\cdot\vec{ \cal   J} + [(\vec E\cdot\vec {\cal  K}+\ii \vec
B\cdot\vec{ \cal   J} ),\del \vec \eta\cdot\vec { \cal K}]
.\]
For the ${\cal K}$'s and $\cal J$'s coefficients we have
\bean
\vec E'=\vec E+\del\vec\eta \times \vec B\el
\vec B'=\vec B-\del\vec\eta \times \vec E ;
\eean
these coupled equations can be written in one, using a complexified
electromagnetic vector field:
\bel{em4}
(\vec E+\ii\vec B)'=(1-\ii\del \vec\eta\times)(\vec E+\ii\vec B),
\ee
corresponding to an infinitesimal imaginary rotation of the quantity
 $\vec E+\ii\vec B$. The finite transformation is therefore
\bel{em5}
(\vec E+\ii\vec B)'=\e^{-\ii\vec\eta\times}(\vec E+\ii\vec B)
\ee
which can be expanded as  in \rf{r10}. Taking the real and imaginary parts
we finally obtain
\bea\label{em6}
\vec E'=\hat{ u}\hat{ u}\cdot \vec E+\sinh\eta\; \hat{ u}
\times\vec B-\cosh\eta(\hat{ u}\times)^2 \vec E\eln
\vec B'=\hat{ u}\hat{ u}\cdot \vec B-\sinh\eta\; \hat{ u}
\times\vec E-\cosh\eta(\hat{ u}\times)^2 \vec B
\eea
which correspond to the usual electromagnetic boost transformations.

We now have that  the square of  transformation \rf{em5} gives
\bel{em7}
(\vec E'+\ii\vec B')^2=(\vec E+\ii\vec B)^2
\ee
{\it i.e.} $E^2- B^2$ and $\vec B\cdot\vec E$ are invariant quantities.
So, if $\vec B\cdot\vec E\neq 0$, the electric an magnetic fields will exist
simultaneously in all inertial frames, while  the angle between the fields stays
 acute or obtuse depending on its value in the original coordinate
frame. 

In the case in which the fields are ortogonal ($\vec B\cdot\vec E=0$), 
it is possible
to find an inertial frame where 

\[ \vec E'=0 \;\;\;{\rm if }\;\;\; B^2>E^2,\;\;\; {\rm
or }\;\;\; 
\vec B'=0 \;\;\;{\rm if}\;\;\; E^2>B^2. \]
Let us clarify this with an example. Consider a particle 
moving in an electromagnetic field 
where $\vec E\cdot\vec B=0$ and $B^2>E^2$ (the case where  $B^2<E^2$ can be 
obtained from this by duality). As we saw, there is  
an inertial
system where  the particle 
is afected only by a  magnetic field $\vec B'$. 
 Using the condition $\vec E'=0$ in the first expresion of \rf{em6} and 
taking both the parallel and perpendicular components with respect  to 
$\hat u$  we find 

\[\hat u\cdot  \vec E =0,\]
\[\sinh\eta\; \hat{ u}
\times\vec B=\cosh\eta(\hat{ u}\times)^2 \vec E;
\]
from which we obtain
\bel{em9}
-\vec{ u}\times\vec B=\vec E,
\ee
where we have used $\vec E=-(\hat{ u}\times)^2 \vec E$  and  $\tanh\eta =u$. 
This equation does not univocally 
determine  $\vec u$, so there are many system where 
the electric field vanishes.
 
In particular we can  
choose the velocity to be ortogonal to the magnetic field,
 obtaining the following 
expresion for the velocity
\bel{em10}
\vec{ u}
=\frac{E}{B}\hat u.
\ee
Because the equation \rf{em5}    corresponds to a rotation, we see that 
the parallel component to $\vec u$ of the electromagnetic field is an invariant, so 
for our case  $\vec B$ and $\vec B'$ must be parallel.
Furthermore, by the invariance of $E^2-B^2$ we  obtain
\bel{em11} \vec B'=\frac{\sqrt{B^2-E^2}}{B}\vec B.\ee 

In this example we saw the utility of the relation \rf{em7} which is 
easilly derived from \rf{em5} and is 
not evident from the usual transformations  \rf{em6}.   
(Usually is derived using tensorial
notation.)

Another interesting example of  Lorentz transformations of the electromanetic
field is when we consider the  evolution of the spin of  a charged particle,   
moving  in a region with an electric 
field  $\vec E$. In the  system in which  the particle is at rest a magnetic field
appears. Its value is  given by the second expresion in \rf{em6} which, in 
the non relativistic  aproximation, is written as 
\[\vec B' =-\vec u\times \vec E.\]
The evolution of the spin is given by
\rf{tp4}  and  \rf{tp3} where $\dot{\vec u} = e\vec E/m$ therefore   
\bel{em12}
\der{\vec s}{t}=
-(g-1)\frac{e}{2m}\vec s\times (\vec u \times \vec E)
\ee 
which is the Thomas equation \cite{5} with $B=0$ and $\gam\ra 1$
As it is well known, this equation gives the correct 
spin-orbit correction in the non relativistic aproximation \cite{6}. 

\section{Conclusions}

We have introduced a way of writing the coordinates of a rotated vector and
deduced the Coriolis acceleration in a straightforward way. The generators
of the rotation group are given a compact form.  In the same spirit we have
obtained Lorentz transformations for 4-vectors and show how the Thomas
precession appears in a non-inertial system after the introduction the
covariant time derivative.
     
Using a matrix construction  we write the non-homogeneous Maxwell equations
in a compact form and, starting from this,  we  deduce the Lorentz
transformations of the electromagnetic fields using the notation introduced
before. We show that the Lorentz transformation of the electromagnetic fields 
can be seen as a
rotations of the  complexified electromagnetic vector $\vec E+\ii\vec B$.
\section*{Appendix: Non-Inertial System}
In this appendix we will  explitly find  the time covariant derivative  
given in \rf{tp2} for non-inertial system. 
We can express this  derivative written 
in terms of the $\cal N$'s,  defined in \rf{s6}, as  
\bel{a1} D_{t}=
\e^{-\frac{1}{2} \vec \eta\cdot\vec {\cal N}} \der {}{t'} 
\e^{\frac{1}{2} \vec \eta\cdot\vec {\cal N}} + {\rm c.c.},
\ee
where we have used the fact that ${\cal N}_i$ and ${\cal N}_j^*$ commute
(Eq.\rf{s8}).
The $\cal N$'s  satisfy the simple relation  
\[ 
{\cal N}_i{\cal N}_i=\del_{ij}+\ii\eps_{ijk}{\cal N}_k,
\]
so  we have
\[ \e^{\frac{1}{2} \vec \eta\cdot\vec {\cal N}} =\cosh
\frac{\eta}{2}+\hat{\eta}.\vec{\cal N}\sinh\frac{\eta}{2}
\]
and therefore
\bel{a2}
\e^{-\frac{1}{2} \vec \eta\cdot{\cal N}} \der {}{t'} \e^{\frac{1}{2} \vec
\eta\cdot{\cal N}} =
\der{}{t'}+\frac{1}{2}\lb \hat{ \eta}\der{\eta}{t'}+
\sinh\eta\der{\hat{\eta}}{t'}-\ii
(\cosh\eta-1)\hat{ \eta}\times\der{\hat{\eta}}{t'}\rb\cdot \vec{\cal N} .
\ee
Finally, returning to  the $\cal J$'s and $\cal K$'s we write
\bel{a3}
D_{t}=\frac{1}{\gam}\der{}{t}+\frac{1}{\gam}\lb\der{\eta}{t}\hat{\eta}+
\sinh\eta\der{\hat{\eta}}{t}\rb \cdot \vec{\cal K}
-\frac{\ii}{\gam}(\cosh\eta-1)\hat{\eta}\times\der{\hat{
\eta}}{t}\cdot \vec{\cal J}
. \ee
In a non-relativistic approximation, $\gam\ra 1$, we have 
 \bel{a4}
D_{t}=\der{}{t}+\frac{\dot{\vec u}}{c}
\cdot \vec{\cal K}-\ii\frac{\vec u
\times\dot{ \vec u}}{2c^2}\cdot \vec {\cal J} . \ee
Considering the covariant derivative acting only on 3-vectors
and  using the definitions of the $\cal J$'s given in \rf{so4}  we obtain   

\bel{a5}
D_{t}=\der{}{t}+\lb\frac{\vec u \times\dot{ \vec u}}{2c^2}\rb\times 
\ee
which is the result \rf{tp4}.

 \end{document}